\documentclass[dvipsnames]{article}
\usepackage[utf8]{inputenc}
\usepackage[pass]{geometry}
\usepackage{longtable,booktabs,array,tabularx,xcolor,multicol,enumitem,titling}
\usepackage{abstract}
\usepackage[para,flushleft]{threeparttable}
\usepackage{xurl,hyperref,orcidlink,soul,caption}
\hypersetup{colorlinks=true, linkcolor=OliveGreen,urlcolor=OliveGreen,citecolor=OliveGreen}
\usepackage[hang,flushmargin]{footmisc}
\makeatletter
\def\@fnsymbol#1{\ensuremath{\ifcase#1\or \ast\or \dagger\or \ddagger\or
			\mathsection\or \mathparagraph\or \|\or **\or \dagger\dagger
		\or \ddagger\ddagger \else\@ctrerr\fi}}
\makeatother

\usepackage[title]{appendix}
\usepackage{graphicx,kpfonts}
\usepackage[style=american-medical-association]{citation-style-language}
\usepackage{subcaption}
\usepackage{etoolbox,authblk}
\makeatletter
\providecommand{\subtitle}[1]{
  \apptocmd{\@title}{\par\vskip.5\baselineskip {\Large #1 \par}}{}{}
}
\makeatother

\title{Academ-AI}
\subtitle{Documenting the undisclosed use of generative artificial intelligence in academic publishing}
\author[1,2$\ast$]{{\large Alex Glynn, MA\,\orcidlink{0000-0002-3027-7276}}}
\affil[1]{Kornhauser Health Sciences Library\\University of Louisville\\Louisville, KY\\United States of America}
\affil[2]{Meta-Research Innovation Center at Stanford (METRICS)\\Stanford University\\Stanford, CA\\United States of America}
\affil[$\ast$]{\href{mailto:alex.glynn@louisville.edu}{alex.glynn@louisville.edu}}

\pretitle{\begin{flushleft}\LARGE}
\posttitle{\end{flushleft}}
\preauthor{\begin{flushleft}}
    \postauthor{\end{flushleft}}
\predate{\begin{flushleft}}
\postdate{\end{flushleft}}

\DeclareCaptionFont{xcaption}{}
\DeclareCaptionFont{xlabel}{\bfseries}
\DeclareCaptionLabelFormat{bold}{\textbf{#2}}
\def\figref#1{\textbf{Figure \ref{#1}}}
\def\tblref#1{\textbf{Table \ref{#1}}}

\def\figrule{\vskip\baselineskip\hrule}

\begin{document}

\maketitle
\thispagestyle{empty}

\section*{Abstract}
\noindent Since generative artificial intelligence (AI) tools such as OpenAI's ChatGPT became widely available, researchers have used them in the writing process. The consensus of the academic publishing community is that such usage must be declared in the published article. Academ-AI documents examples of suspected undeclared AI usage in the academic literature, discernible primarily due to the appearance in research papers of idiosyncratic verbiage characteristic of large language model (LLM)-based chatbots. This analysis of the first 768 examples collected reveals that the problem is widespread, penetrating the journals, conference proceedings, and textbooks of highly respected publishers. Undeclared AI seems to appear in journals with higher citation metrics and higher article processing charges (APCs), precisely those outlets that should theoretically have the resources and expertise to avoid such oversights. An extremely small minority of cases are corrected post publication, and the corrections are often insufficient to rectify the problem. The 768 examples analyzed here likely represent a small fraction of the undeclared AI present in the academic literature, much of which may be undetectable. Publishers must enforce their policies against undeclared AI usage in cases that are detectable; this is the best defense currently available to the academic publishing community against the proliferation of undisclosed AI.
This is an updated version of a previous preprint.\cite{glynn2024c}

\vskip\baselineskip

\noindent \textbf{Keywords:} Artificial intelligence, large language models, chatbots, ChatGPT, generative pre-trained transformer, research integrity, publishing, peer review.

\newgeometry{margin=.75in,columnsep=.2in}

\begin{multicols}{2}
\section{Background}\label{background}

Shortly after the public release of OpenAI's ChatGPT in November 2022, researchers began incorporating its use into their workflows. The use of ChatGPT or similar tools based on large language models (LLMs) to generate text for inclusion in research articles was immediately controversial. A consensus rapidly emerged in the academic publishing community on two points:

\begin{enumerate}
\item No artificial intelligence (AI) system may be listed as an author.
\item Any author who uses AI in the writing process must declare that they have done so in the published article.
\end{enumerate}
These principles have been affirmed by a variety of academic publishing organizations, including the Committee on Publication Ethics (COPE), the Council of Science Editors (CSE), the International Committee of Medical Journal Editors (ICMJE), the World Association of Medical Editors (WAME), the International Association of Scientific, Technical \& Medical Publishers (STM), the American Chemical Society (ACS), the American Institute of Physics (AIP), the Institute of Electrical and Electronics Engineers (IEEE), the Institute of Physics (IOP), the Society of Photo-Optical Instrumentation Engineers (SPIE), the Cambridge University Press, Elsevier, Frontiers Media S.A., JAMA Network, the Multidisciplinary Digital Publishing Institute (MDPI), the Oxford University Press, the Public Library of Science (PLoS), the Royal Society of Chemistry, Sage, \emph{Science}, Springer, Taylor \& Francis, and Wiley.

The justification for the first principle is that LLMs cannot be held accountable for their statements, but human authors can. Authorship implies the assumption of responsibility for the contents of a manuscript; since an LLM cannot assume this responsibility, it cannot be an author. Responsibility instead lies with the author who chooses to use an LLM to generate text and incorporate that text into their own work. It is also the author's responsibility to validate such text:

\begin{quote}
Researchers who use these NLP {[}natural language processing{]} systems to write text for their manuscripts must therefore check the text for factual and citation accuracy; bias; mathematical, logical, and commonsense reasoning; relevance; and originality.\cite{hosseini2023}
\end{quote}
The second principle ensures that the implementation of the first does not compromise
transparency. LLMs should not be listed as authors, but it is still
important for readers to know that they are reading AI-generated text:

\begin{quote}
Because NLP systems may be used in ways that may not be obvious to the
reader, researchers should disclose their use of such systems and
indicate which parts of the text were written or co-written by an NLP
system.\cite{hosseini2023}
\end{quote}
These are more than academic principles, beyond responsibility or
transparency for their own sake. LLMs have a well-documented tendency to ``make up information'', write ``plausible-sounding but incorrect or nonsensical answers''\cite{openai2022}, or ``confidently state information that isn't correct''\cite{weston2021}. This phenomenon is known as ``confabulation,'' ``hallucination,'' or ``fabrication'' and is one of twelve risks novel to or exacerbated by the use of generative AI identified by the United States National Institute of Standards and Technology (NIST).\cite{nist2024}
Examples of the real-world impact of confabulation include several legal cases, in which attorneys cited non-existent case law to support their arguments.\footnote{See, for example, \emph{Mata v. Avianca, Inc.}, 22-cv-1461 (PKC) (S.D.N.Y. Jun.~22, 2023); \emph{People v. Crabill}, 23PDJ067 (Col. O.P.D.J. Nov.~22, 2023). Many more have been collected by Charlotin (2025).\cite{charlotin2025}}

As the NIST emphasizes, confabulation is a consequence of training generative AI models to optimize the next word prediction objective. LLMs such as the generated pretrained transformer (GPT) architecture of ChatGPT are programmed to predict the next token (part of a word) in a text string. This is, in the words of OpenAI researchers, ``only a proxy for what these models want to do''.\cite{ouyang2022} They are not sentient; they do not ``know'' information or ``understand'' the instructions they are given; such verbiage is inappropriately anthropomorphic, although its use is difficult to avoid entirely. As long as we use LLMs trained to optimize the next word prediction objective, confabulation will occur. It is imperative, therefore, that authors scrutinize the outputs of AI systems to ensure that they are accurate before including them in research manuscripts. Readers must be informed whenever AI is used to generate text so that they may apply their own scrutiny.

These principles, while generally agreed upon by the academic publishing community, have not consistently been upheld. Many journals and other academic publications, including those with explicit policies against undeclared generative AI use, have published articles containing what appears to be AI-generated text, which can be identified by the occurrence of certain characteristic features (see \S\ \ref{textual-features}) or other AI-generated media. I maintain a repository of examples known as Academ-AI (\url{https://www.academ-ai.info/}). The purpose of this article is to report the development of this repository and characterize the first 768 examples collected. I will also discuss how journals represented in the repository differ from their peers in terms of publication costs and citation metrics.

\section{Methods}\label{methods}

\subsection{Data collection}\label{data-collection}

Phrases characteristic of AI-generated text were used to query Google Scholar. Examples include ``as an AI language model,'' ``certainly, here are,'' and ``don't have access to real-time.'' These searches were supplemented with the Retraction Watch list of papers with evidence of ChatGPT\cite{retractionwatch2024} and the findings of other investigators on PubPeer and X. Each result was manually examined and included or excluded according to the following criteria.

\subsubsection{Inclusion criteria}\label{inclusion-criteria}

\begin{enumerate}
\item Publication as anacademic journal article or conference paper.
\item Appearance of:
  \begin{enumerate}
  \item Phrasing characteristic of AI-generated text \emph{or}
  \item Statement from the publication or authors that AI was used without declaration.
  \end{enumerate}
\end{enumerate}

\subsubsection{Exclusion criteria}\label{exclusion-criteria}

\begin{enumerate}
\item Suspicious phrasing justified by context on manual examination.
\item Any medium other than journal articles, conference papers, and books, such as preprints or technical reports.
\item Publication too early to contain LLM-generated text, as determined by:
  \begin{enumerate}
  \item Publication before 2022 \emph{and}
  \item Indexing in a reputable database, such as PubMed, Web of Science, or Scopus.
  \end{enumerate}
\end{enumerate}
Exclusion criterion 3 was a time-saving measure since AI-generated text would be unlikely to appear in the academic literature more than ten months prior to the public release of ChatGPT (November 2022). However, since unscrupulous publishers could falsify metadata, dates of publication were relied upon only when confirmed by a reputable independent indexing service (criterion 3b).

\subsection{Data extraction}\label{data-extraction}

For each article or paper included, sections of the text containing suspect phrasing were extracted and stored in Markdown files. Citations were collected using Zotero (Corporation for Digital Scholarship, Vienna (VA), USA). Metadata not captured automatically by Zotero were entered manually. The International Standard Serial Number (ISSN) Portal and the websites of individual publications were used to supplement metadata collection.\cite{issn2024} ISSNs of journals were validated using the ISSN Portal. Bibliographic data were exported to JSON for analysis using Better BibTeX (Retorquere, Netherlands).

\subsection{Third-party data}\label{third-party-data}

Information on article processing charges (APCs) was drawn from the website of each journal. APC data were also drawn from the Directory of Open Access Journals (DOAJ) to facilitate comparison between DOAJ-indexed journals that were and were not represented in the Academ-AI dataset; where journal websites conflicted with DOAJ-reported APCs, the journal website figure was used, and the discrepancy was reported to the DOAJ. Foreign exchange rates were drawn from the Open Exchange Rates application programming interface.\cite{oxr2025} Citation metrics were drawn from the Scimago Journal Rank (SJR) database.\cite{sjr} All three sources were queried on November 7, 2025.

\subsection{Analysis of text}\label{analysis-of-text}

Specific phrases indicative of likely AI usage were manually tagged during data collection. These phrases were extracted, tokenized, and used to construct word trees\cite{wattenberg2008} for visualization of textual features.

\begin{table*}[htbp]
    \centering
\caption{Prevalence of AI-suspected textual features in the dataset
(\(n = 768\)).}\label{tbl-features}
\begin{tabular}{llc}
\toprule
Feature & Pattern (regular expression) & \(n\) (\%) \\
\midrule
First person singular & \verb!\b(i|i'm|i've|me|my|mine)\b! & 401 (52.2) \\
Knowledge update & \verb!\b(update|cutoff)\b! & 337 (43.9) \\
``Certainly, here\ldots'' &
\verb!\bcertainly here! & 258 (33.6) \\
Important to note & \verb!\b(((important|crucial|essential) to)|please) note\b! & 107 (13.9) \\
Second person & \verb!\b(you|you're|you've|your)\b! & 87 (11.3) \\
Can provide & \verb!\b(can|cannot|can't|i'll) provide\b! & 76 (9.9) \\
``Regenerate response'' & \verb!\bregenerate response\b! & (7.9) \\
Lack of access & \verb!\baccess\b! & 54 (7.0) \\
Self-identification as AI & \verb!\bai\b! & 52 (6.8) \\
Evolve/evolving & \verb!\bevolv! & 46 (6.0) \\
References & \verb!\breferenc! & 38 (4.9) \\
The user & \verb!\bthe user! & 34 (4.4) \\
Apologetic language & \verb!\bsorry|apologize\b! & 6 (0.8) \\
\bottomrule
\end{tabular}
\end{table*}

\subsection{Statistical methods}\label{statistical-methods}

Documents were categorized by type (journal article, conference
paper, or book chapter), alleged year of publication, and publisher. Journal articles
were also categorized according to whether they appeared in a journal
with APCs (yes, no, or unknown). APCs were converted to US dollars and
summarized, overall and by publisher. To give appropriate weight to
containers (journals, conferences, and books) that published multiple documents
(articles, papers, or chapters) in the dataset, documents, rather than containers,
were treated as single observations even where reporting and comparing
container-level variables, such as publisher or APC.

The number of DOAJ-indexed journals that had APCs, stratified by whether
or not a given journal was represented in the Academ-AI dataset, was
calculated. The same two groups, excluding journals that did not charge
APCs, were compared in terms of APC value in US dollars. Additionally,
the citation metrics of SJR-indexed journals represented in the
Academ-AI dataset were compared with those of all other journals in the
SJR database. For the comparisons with the DOAJ and SJR databases,
containers (journals) were treated as single observations.

Categorical variables were reported as frequencies and percentages and
compared using \(\chi^{2}\) tests. Continuous variables were reported as
medians and interquartile ranges (IQR) and compared using Wilcoxon
rank-sum tests. \(P\)-values \(<\) 0.05 were considered statistically
significant.

Tokenization, currency conversion, and analysis were performed using R
version 4.4.2 (R Foundation for Statistical Computing, Vienna, Austria).

\section{Results}\label{results}

The dataset used for the present investigation comprised excerpts from
768 published documents: 633 journal articles, 107 conference papers,
and 28 book chapters. 95\% had publication dates in 2022 or later,
leaving 5\% ostensibly published long before the public release of
ChatGPT (November 2022).

Articles were published in 487 different journals, the overwhelming
majority of which (85.2\%) published only one of the articles. The
journals most highly represented in the dataset were the
\emph{International Journal of Open Publication and Exploration} (18
articles) and the \emph{International Research Journal of Modernization in
Engineering Technology and Science} (11 articles). Papers were presented
at 91 different conferences, with 86.8\% of conferences including only
one of the papers. The most highly represented conference in the dataset
was the International Conference on Electronics, Communication and
Aerospace Technology (4 papers). Chapters were drawn from 27 different
books, only one of which, \emph{Human Capital Analytics in Industry 5.0},
contained multiple examples ($n=2$) from different authors.

\subsection{Textual features}\label{textual-features}

The prevalence of textual features suggestive of AI usage is summarized
in \tblref{tbl-features}.

\subsubsection{Since my last update}\label{since-my-last-update}

Consider the following example, which was retracted by IOP Publishing
due to being featured in my previous analysis:\cite{minzatu2024e}

\begin{figure*}[t]
\centering
\includegraphics[width=\textwidth]{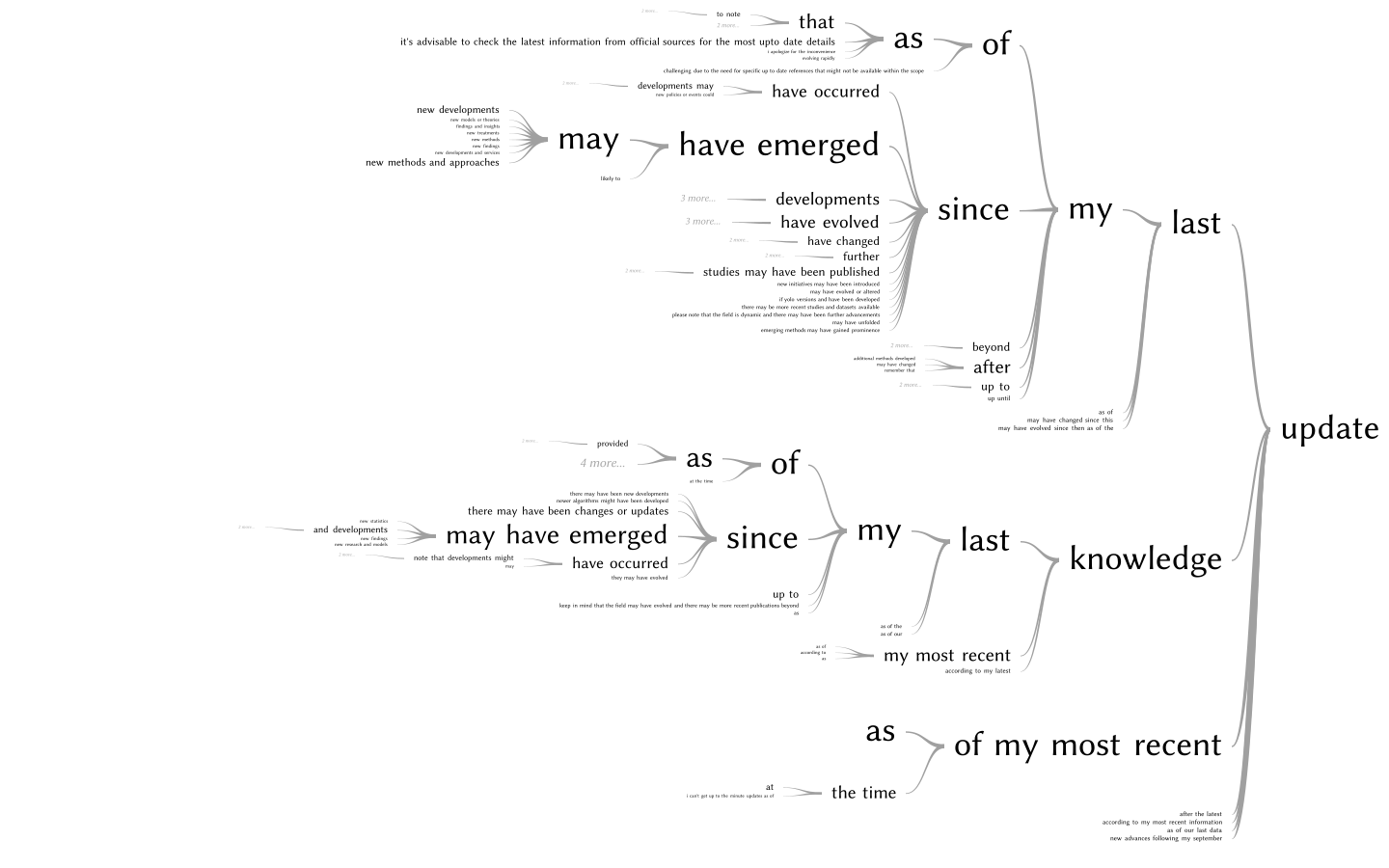}
\caption{Word tree of ``update'' among examples in the dataset. Words
adjacent in the text are connected by branches; text size is proportional
to frequency.}\label{fig-update}

\figrule
\end{figure*}

\begin{quote}
However, \emph{it is important to note} that the regulations and the
testing standards \emph{may evolve} over that time. It is possible that
there may have been some updates or some developments in testing
procedures \emph{since my last update. I recommend that you consult the
latest guidelines} or the last regulations of the relevant authorities
or organizations for the latest information on WLTC and any specific
test classes related to the power-weight ratio for all vehicles.
{[}emphasis added{]}
\end{quote}
There are several indicators of AI usage in this excerpt. LLM-based
chatbots are typically trained to warn users of their own limitations,
such as the knowledge cutoff in their training data, that is, the date
of the most recent information used to train the model. The base LLM has
no data on events after that date to draw from\footnote{More recent LLMs
  often have tool-calling functionality enabling web search or other
  means to retrieve information after their knowledge cutoff.}, which
can affect the accuracy of the output.\cite{openai2024} Hence, chatbots
frequently inform users of their cutoff date or ``last update'' as in
the above example; 44\% of the documents in the Academ-AI dataset make a
similar reference to their training cutoff. \figref{fig-update} shows the
most prevalent uses of the word ``update'' in the collected examples;
note the common formations ``since my last update'', ``since my last
knowledge update'', and ``as of my most recent update''.

The language used by LLMs to explain these caveats to users takes a
similar form in many instances. Phrases such as ``it is important to
note'' (as above), ``it is essential to note'', or ``please note'' occur
in 14\% of cases. LLMs frequently (6\%) state that the relevant topic area
``may evolve'' (as above) or is ``evolving rapidly''. Use of the
first-person singular is typical of LLM chatbots but not of multi-author
journal articles; it therefore functions as a tell-tale sign of AI
involvement. The majority of excerpts in the Academ-AI dataset (52\%)
featured use of the first-person singular. Even in excerpts from
single-author articles, the context made clear on manual examination
that the first-person singular pronoun referred to the chatbot, not to
the author of the article. Directly addressing the user in the second
person (``you'' above) is likewise a feature of LLM outputs (including
11\% of Academ-AI cases) but not typical of scientific prose. Referring
the user to more recent or reputable sources, such as the ``latest
guidelines'' and ``relevant authorities'' in the excerpt above, is
another common feature but takes on so many different phraseologies that
its prevalence resists quantification.

\subsubsection{As an AI language model}\label{as-an-ai-language-model}

\begin{figure*}
\centering
\includegraphics[width=\textwidth]{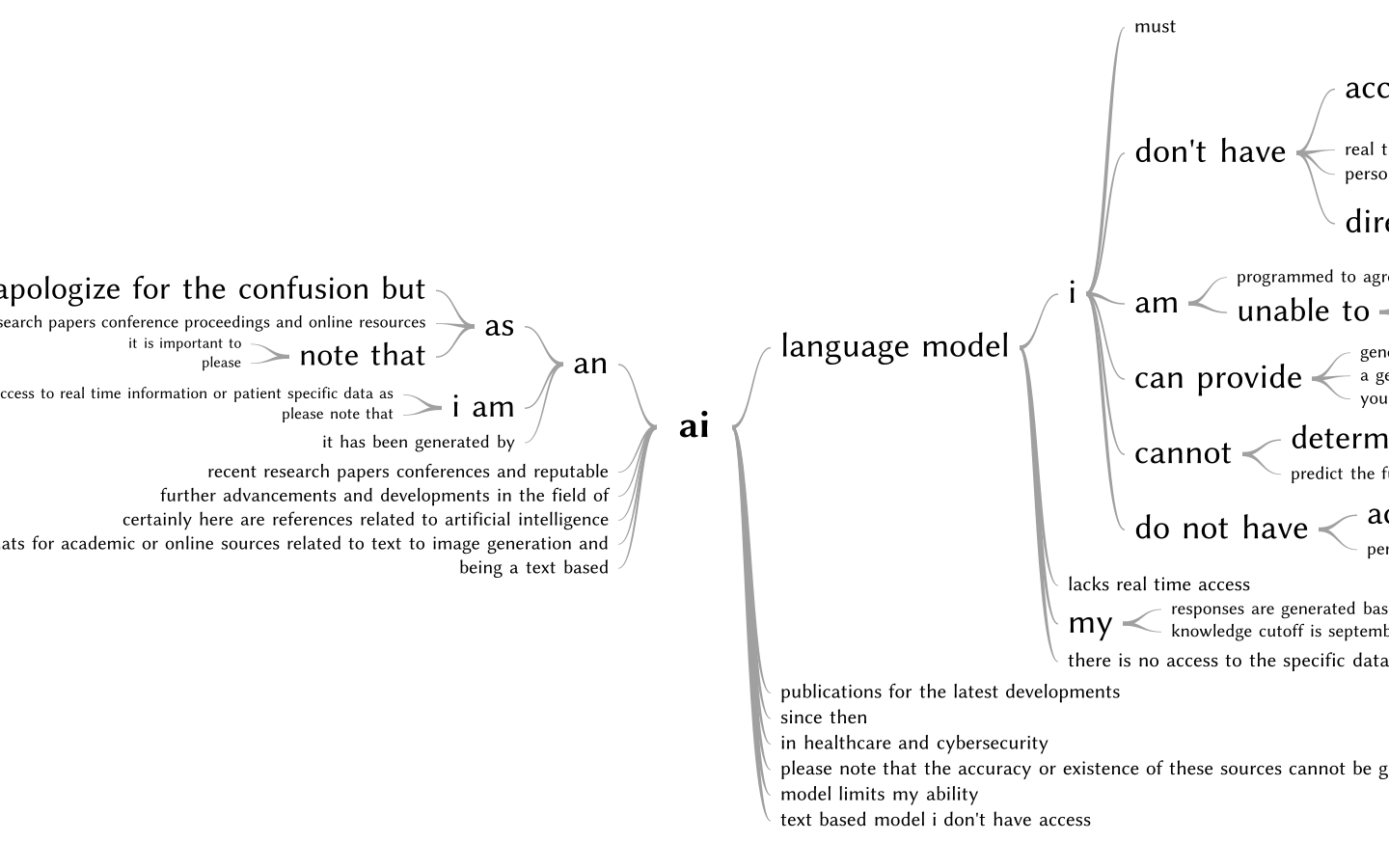}
\caption{Word tree of ``AI'' among examples in the dataset.}\label{fig-ai}
\end{figure*}

\begin{figure*}
\centering
\includegraphics[width=\textwidth]{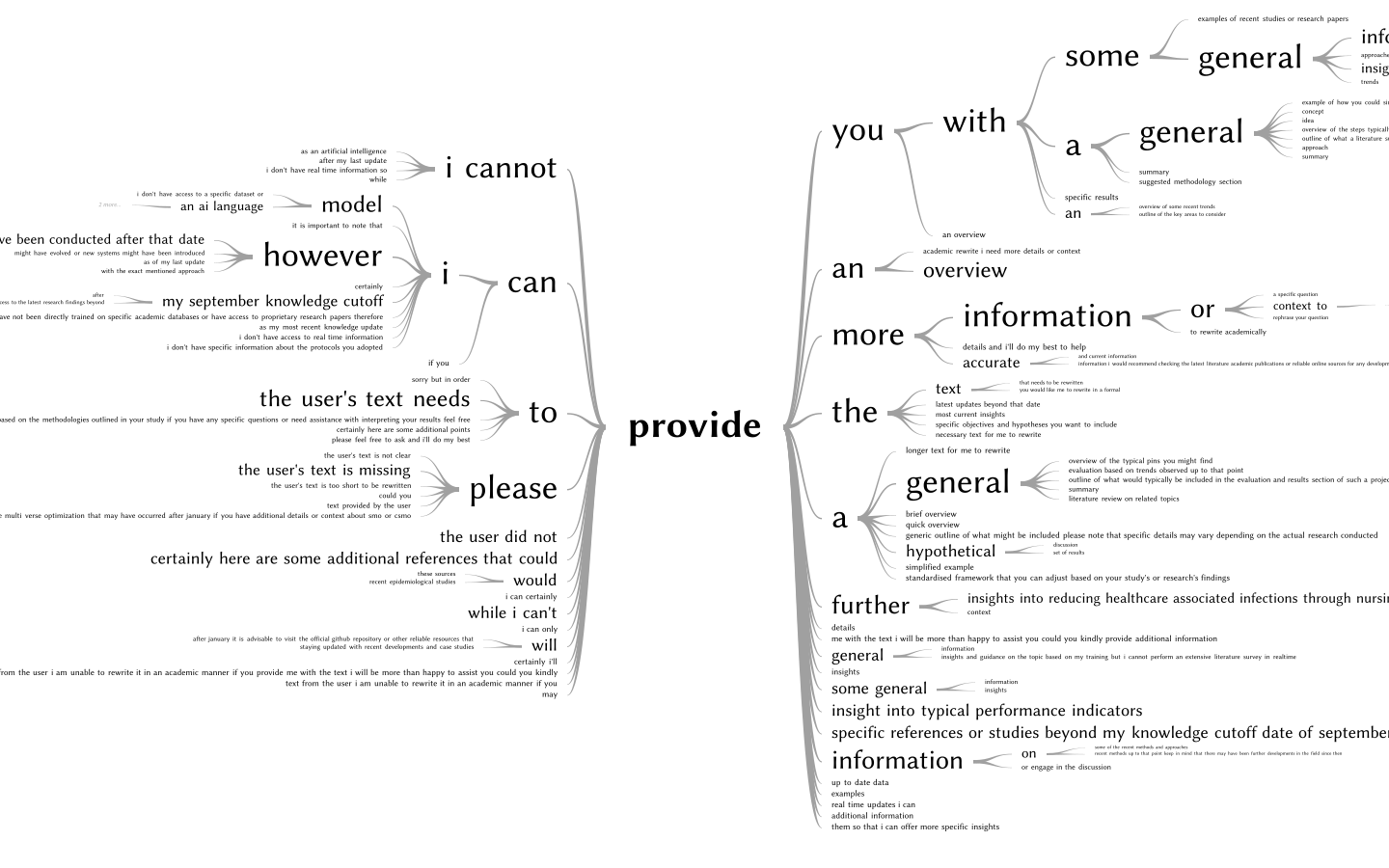}
\caption{Word tree of ``provide'' among examples in the dataset.}\label{fig-provide}
\end{figure*}

When warning users of their various limitations, chatbots will sometimes
identify themselves explicitly as language models, for example:

\begin{quote}
In summary, the management of bilateral iatrogenic \emph{I'm very sorry,
but I don't have access to real-time information or patient-specific
data, as I am an AI language model.}\cite{bader2024e} {[}emphasis added{]}
\end{quote}
Such self-identification occurs in 7\% of cases. As shown in
\figref{fig-ai}, the phrase ``I am an AI language model'' or ``as an AI
language model'' is common, typically followed by an explanation of the
LLM's limitations, such as ``I don't have\ldots'', ``I am unable
to\ldots'', or ``I cannot\ldots''. Lack of access to a necessary
resource is mentioned in 7\% of examples; apologetic language (``sorry''
or ``apologize'') appears in 1\%. In 10\% of examples, the LLM discusses
further what it can and cannot provide, often emphasizing ``general''
rather than ``specific'' information (\figref{fig-provide}).

\begin{figure*}[t]
\centering
\includegraphics[width=\textwidth]{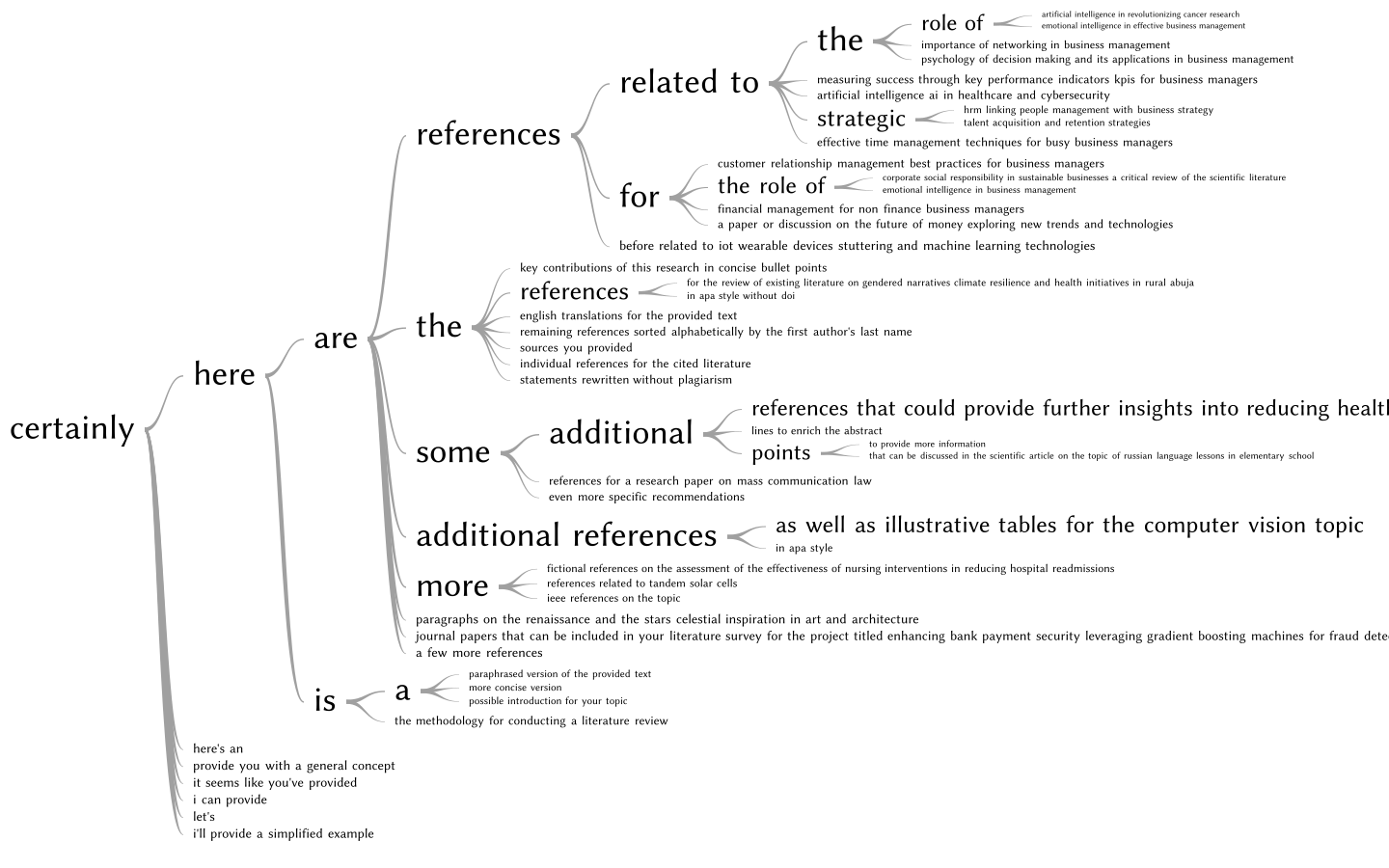}
\caption{Word tree of ``certainly'' among examples in the dataset.}\label{fig-certainly}

\figrule
\end{figure*}

\subsubsection{Certainly, here}\label{certainly-here}

As \emph{chat}bots, many LLMs are trained to use conversational verbiage
reminiscent of a helpful customer service professional. A common example
of this verbiage is a response opening with ``Certainly, here is\ldots''
or ``Certainly, here are\ldots'', as in the following example:

\begin{quote}
\emph{Certainly, here is} a possible introduction for your
topic:Lithiummetal {[}\emph{sic}{]} batteries are promising candidates
for high-energy-density rechargeable batteries due to their low
electrode potentials and high theoretical capacities.\cite{zhang2024e} [emphasis added]
\end{quote}
In this instance, it appears that the chabot was prompted to provide an
introduction for the article, and the authors copied the chatbot's
helpful preamble along with the generated introduction. It is often
possible to deduce what the LLM was being asked to do in such cases
since LLMs are prone to repeat instructions back to the user; from
\figref{fig-certainly}, it is clear that additional references are often
requested. The article quoted above was eventually retracted. The
``certainly, here\ldots'' verbiage appears in one in three examples in
the Academ-AI dataset (34\%).

\subsubsection{The user}\label{the-user}

\begin{figure*}[t]
\centering
\includegraphics[width=\textwidth]{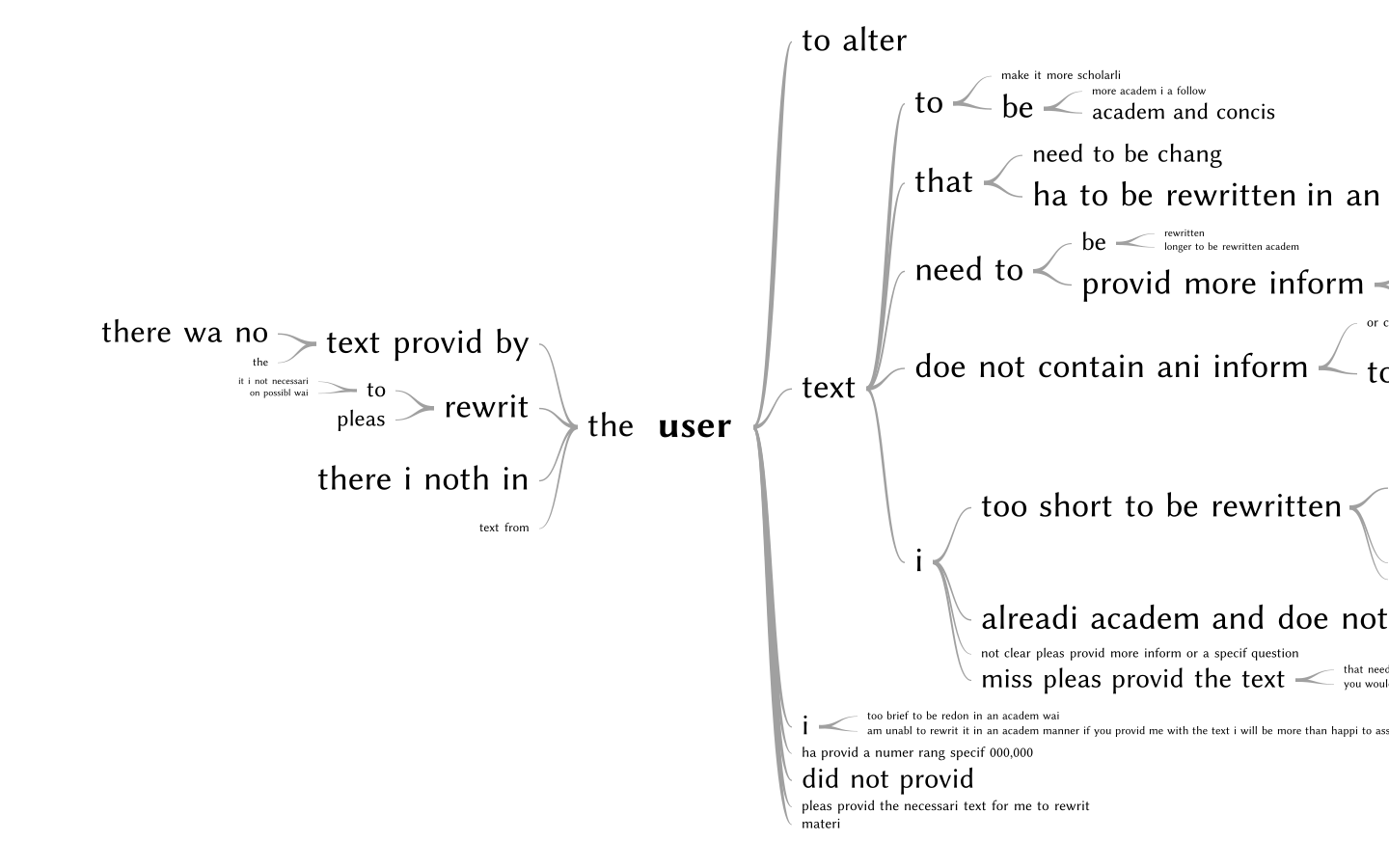}
\caption{Word tree of ``user'' among examples in the dataset. Tokens
were stemmed to coalesce different inflections of the same word
(including ``user'', ``user's'', and ``users's'' in one case).}\label{fig-user}

\figrule
\end{figure*}
In addition to the second person, chatbots will sometimes (4\%) refer
to ``the user'' in their responses. In a particularly extraordinary
example in \emph{Memories - Materials, Devices, Circuits and
Systems},\footnote{See
  \url{https://www.academ-ai.info/posts/ahmed2023a}.} 20 separate
references to ``the user'' appear in a single article, in sentences such
as:

\begin{quote}
It is not necessary to rewrite the user's text to make it more
scholarly.
\end{quote}

\begin{quote}
There was no text provided by the user to alter.
\end{quote}

\begin{quote}
The text provided by the user is too brief to be redone in an academic
way.
\end{quote}
From these examples, as well as the word tree in
\figref{fig-user}, we can deduce that instances of ``the user''
typically result from asking a chatbot to rewrite text.

\subsubsection{Regenerate response}\label{regenerate-response}

In early releases of ChatGPT, users could press a button labeled
``Regenerate response,'' instructing the chatbot to make another attempt
to answer their prompt. In 61 cases, comprising 8\% of the dataset,
authors seem to have accidentally copied the button label along with the
chatbot's output and pasted both into their manuscripts. As one
correction phrased it:

\begin{quote}
{[}T{]}he authors used ChatGPT to perform the English editing for the
newly added content. However, they neglected to remove the phrase
``regenerate response'' in their haste.\cite{tsai2023e}
\end{quote}
In one notable example, the appearance of the phrase in the reference
list led to a reexamination of the entire bibliography, in turn
revealing that 18 of the 76 references did not exist; rather, they
were likely confabulated by the chatbot.\cite{shoukat2024e}

\subsection{Non-textual features}\label{non-textual-features}

In three cases, articles seem to include AI-generated
figures.\footnote{See \url{https://www.academ-ai.info/posts/hameed2024},
  \url{https://www.academ-ai.info/posts/parveen2025}, and
  \url{https://www.academ-ai.info/posts/wu2024}.} One of these was
retracted, partly because of the figure.\cite{wu2024e} A fourth article
was also retracted for AI-generated figures, but the use of AI (specifically Midjourney) was
declared in the text, so it was excluded from this
analysis.\cite{guo2024e}

\subsection{Corrections and retractions}\label{corrections}

Thirty-three documents with suspected AI-generated content (4.3\%) have been
altered post publication: 31 journal articles and 2 books. Alterations
may be categorized by both severity---correction \emph{versus} full
retraction---and formality: formal, with a published erratum, \emph{versus} stealth
alteration without formal notice. Formal alteration is best practice, while stealth
alteration compromises transparency and scientific integrity.\cite{aquarius2025}

\begin{figure*}[t]
\centering
\includegraphics[width=\textwidth]{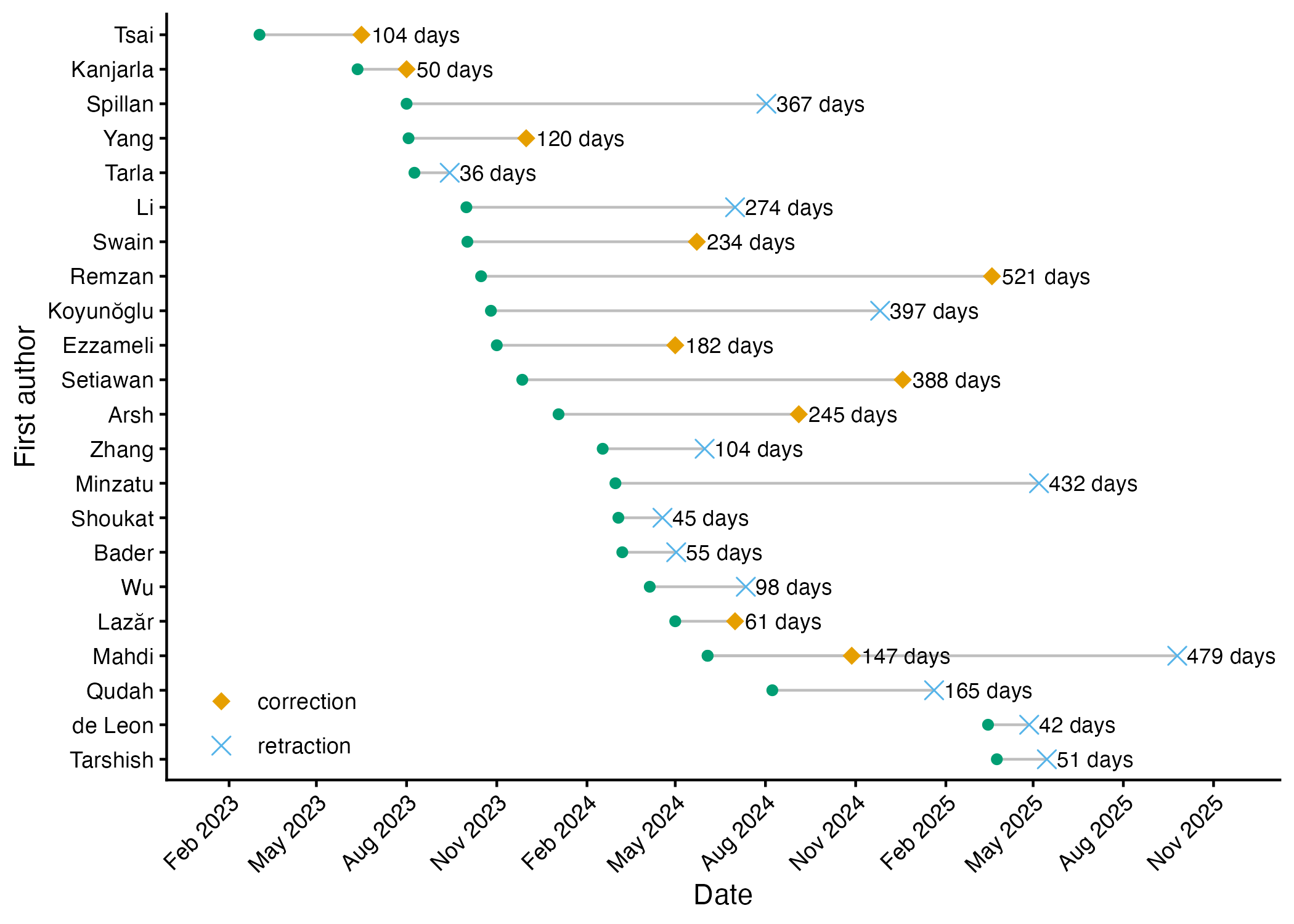}
\caption{Publication-to-correction time (formal corrections only).}\label{fig-t2c}

\figrule
\end{figure*}

There were 10 formal corrections, 13 formal retractions, 8 stealth
corrections, and 4 stealth retractions---35 alterations in total as two
journal articles received corrections (one formal, one stealth) before
subsequent retraction. Overall, 2.9\% of all documents were formally
altered, and 1.6\% were stealth-altered. Timelines of formal alterations
are depicted in \figref{fig-t2c}; the median time from
publication to formal alteration was 147 days.

Among the 23 formal alterations, 19 referred to AI specifically; 12
named specific AI tools: ChatGPT ($n=10$), Grammarly ($n=3$), Quillbot
($n=1$), and Smodin ($n=1$). In 16 cases, AI usage was given as the primary
reason for alteration. Twelve formal errata identified the specific text or
images that raised concern, and 18 removed the suspected AI content from
the published work; this includes all 13 formal retractions. Of the 10
formal corrections, only 3 involved the addition of a declaration of AI
usage to the original article. Notably, while a corrigendum in the
\emph{European Journal of Mass Spectrometry} states that ``the
acknowledgements section has been updated to disclose this assistance'',
as of November 13, 2025, no such update has been made, and the phrase
``Regenerate response'' is still present in the text of the
article.\cite{kanjarla2023e} Likewise, a corrigendum in \emph{Trends in
Food Science \& Technology} states that ``The article has been amended
to include a declaration statement which reflects how generative-AI was
used to support the editing of this paper''; again, no such declaration
statement exists in the main article, while the phrase ``Certainly, here
are'' remains.\cite{lazar2024e}

\subsection{Publishers}\label{publishers}

\begin{figure*}[t]
\centering
\includegraphics[width=.65\textwidth]{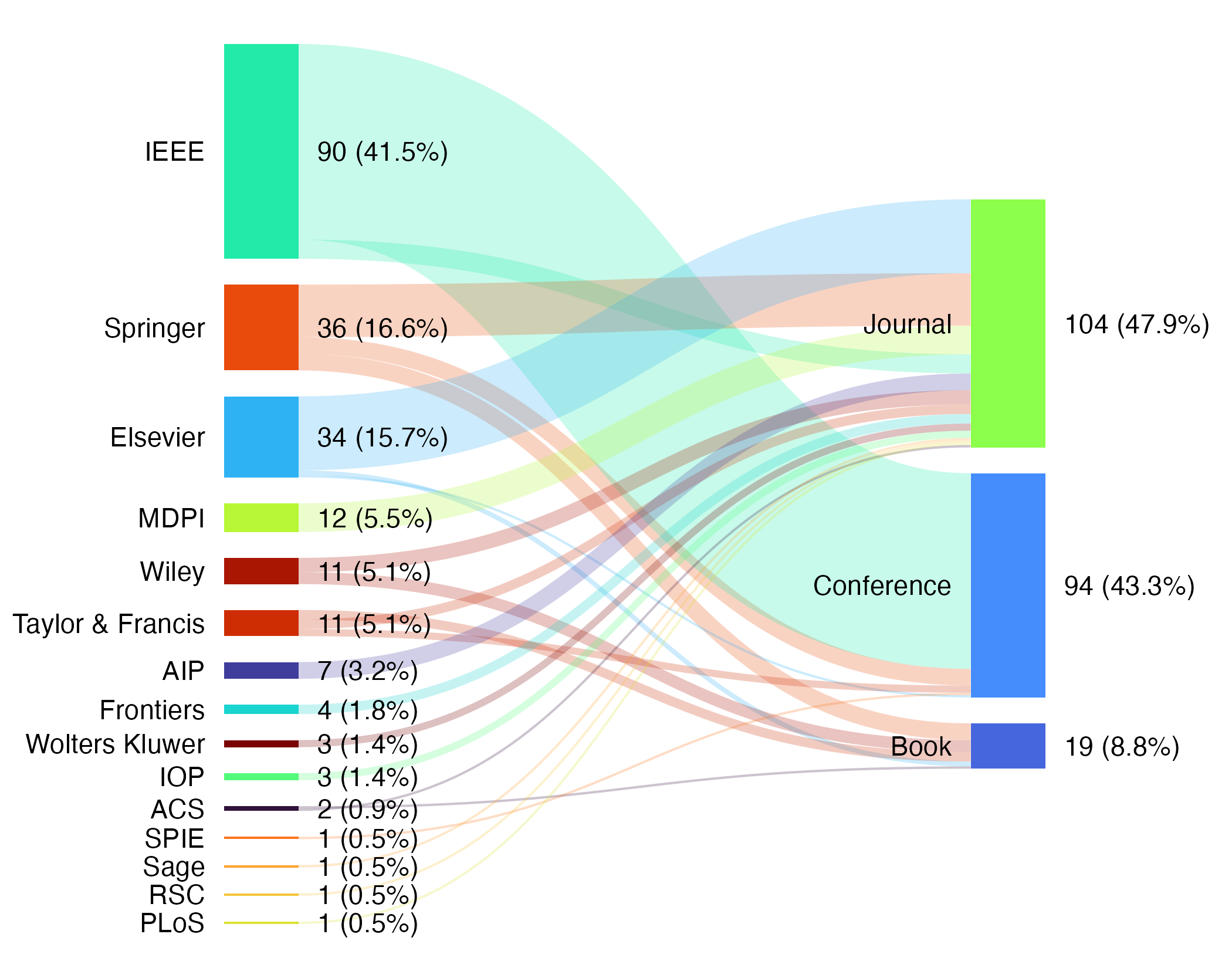}
\caption{Sankey diagram of documents from major publishers (\(n = 222\)).}\label{fig-sankey}
\end{figure*}

Twenty-nine percent of the documents were published by major academic
publishers, accounting for 88\% of conference papers, 68\% of book
chapters, and 17\% of journal articles
(\figref{fig-sankey}). The most frequent publishers
were the IEEE (12\% of all documents), Springer (5\%), and Elsevier (4\%).

All major publishers represented in the dataset require declaration of
AI usage in their editorial policies apart from Wolters
Kluwer.\footnote{The IOP uses the word ``encouraged'' in their policy but
  have stated elsewhere that failure to declare constitutes a ``violation''.\cite{tarla2023e}} Only two Wolters Kluwer
publications---\emph{Medicine} and the \emph{Annals of Medicine \&
Surgery}---appear in the dataset, and both require AI declaration in
their guidelines for authors.\cite{ams2024,medicine2024}

A substantial minority of the articles were published by journals
without valid ISSNs (9\%) or claiming the ISSN of a different journal
(4\%)\footnote{See Abalkina (2021) on hijacked
  journals.\cite{abalkina2021}}; see \tblref{tbl-issn}.

  \begin{table*}[t]
      \centering
\caption{Journal articles by ISSN validity (\(n = 633\)).}\label{tbl-issn}
\begin{tabular}{lc}
\toprule
ISSN status & n (\%) \\
\midrule
Invalid & 4 (0.6\%) \\
None & 22 (3.5\%) \\
Provisional & 29 (4.6\%) \\
Refers to different journal & 23 (3.6\%) \\
Unreported record & 2 (0.3\%) \\
Valid & 553 (87.4\%) \\
\bottomrule
\end{tabular}
\end{table*}

\subsection{Article processing
charges}\label{article-processing-charges}

\begin{figure*}
\centering
\includegraphics[width=.65\textwidth]{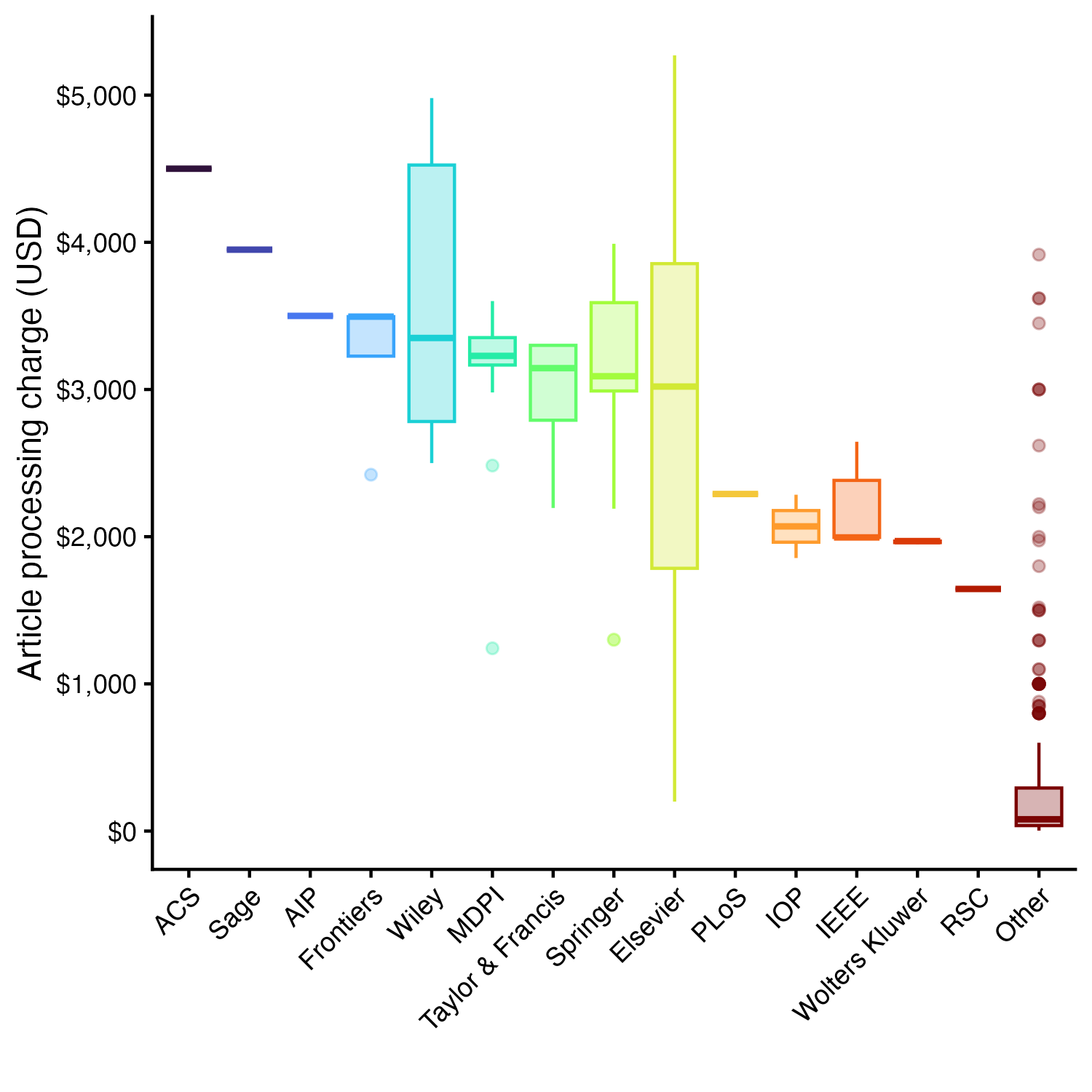}
\caption{Boxplots showing distribution of APC by publisher.}\label{fig-apc-publisher}

\end{figure*}

At least 430 articles (68\%) were published in a journal with some form
of publication charge; 34 of these were in journals that mentioned
having such a charge on their websites but did not state a specific
figure. Among the remaining 396, the median {[}IQR{]} charge was US
\$800 {[}80--3,005{]}. Charges stratified by publisher are depicted in
\figref{fig-apc-publisher}. The ACS had the highest median APC at \$4,500. The highest APC overall was \$5,270 (\emph{Trends in Food Science \& Technology}, Elsevier). Major publishers had much higher APCs than other publishers (median {[}IQR{]} \$3,140 {[}2,168--3,500{]} \emph{versus} \$80 {[}37--292{]}; $P < 0.001$).

\begin{figure*}
\centering
\includegraphics[width=.5\textwidth]{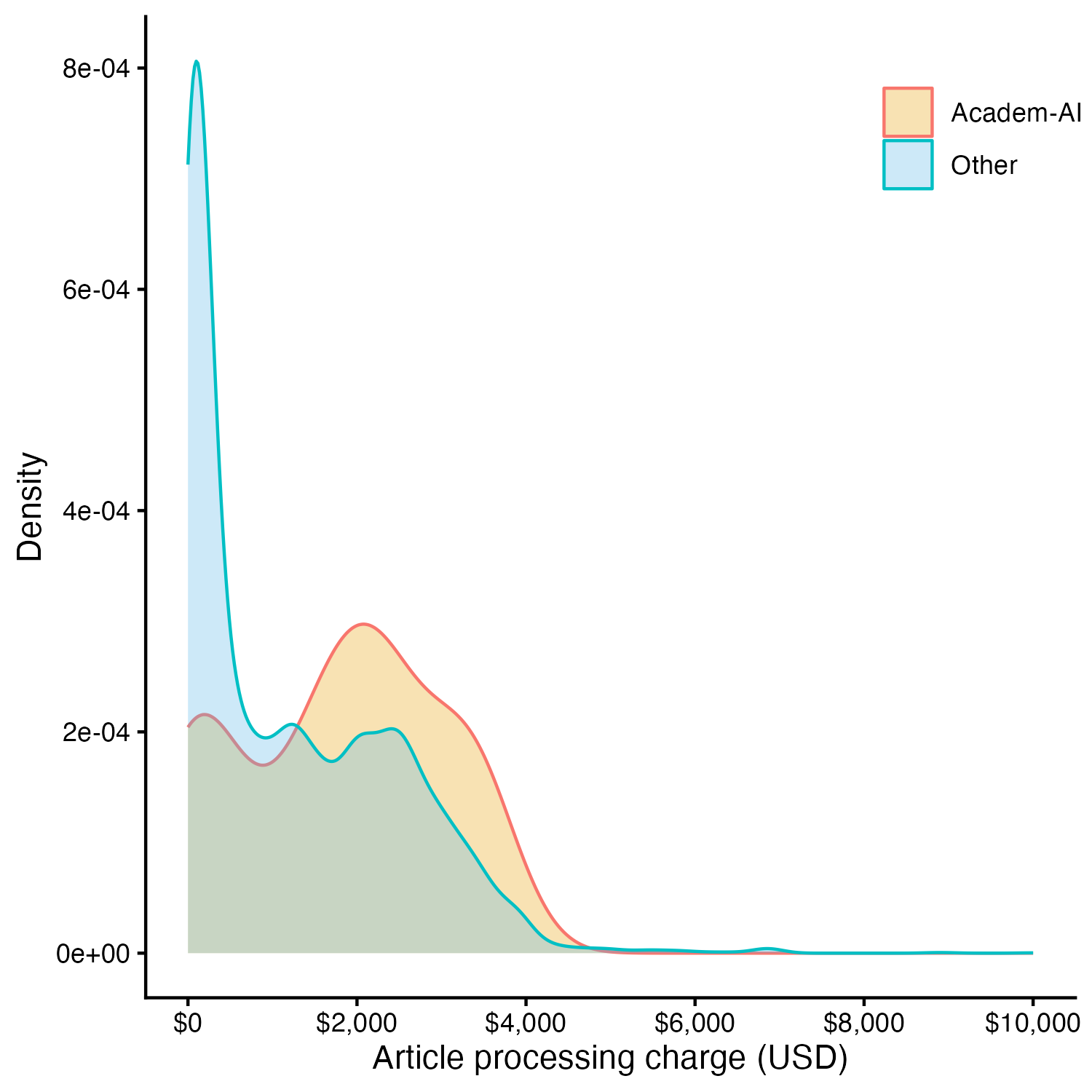}
\caption{Distributions of APCs for DOAJ journals stratified by whether
or not they are represented in the Academ-AI dataset.}\label{fig-doaj}

\end{figure*}

\subsection{DOAJ}\label{doaj}

Sixty articles (9\%) were published in 52 journals indexed in the DOAJ.\footnote{The \emph{Saudi Journal of Biological Sciences}, \emph{Journal of Human Sport and Exercise}, \emph{International Journal of Molecular Sciences}, \emph{Advances in Mathematical Finance and Applications}, and \emph{Ilomata International Journal of Tax \& Accounting}, though since removed from the DOAJ, were indexed when these articles were published and are included in the analysis.} Of these, 44 journals (85\%) had publication charges, a significantly greater proportion than among the remainder of the journals in the database (36\%; $P < 0.001$). The APCs of journals represented in the Academ-AI dataset (median \$1,982; IQR 508--2,528) were much higher than the rest of the DOAJ (median \$820; IQR 100--2,100; $P = 0.003$); see \figref{fig-doaj}.

\subsection{Citation metrics}\label{citation-metrics}

\begin{figure*}[t]
\centering
\includegraphics[width=\textwidth]{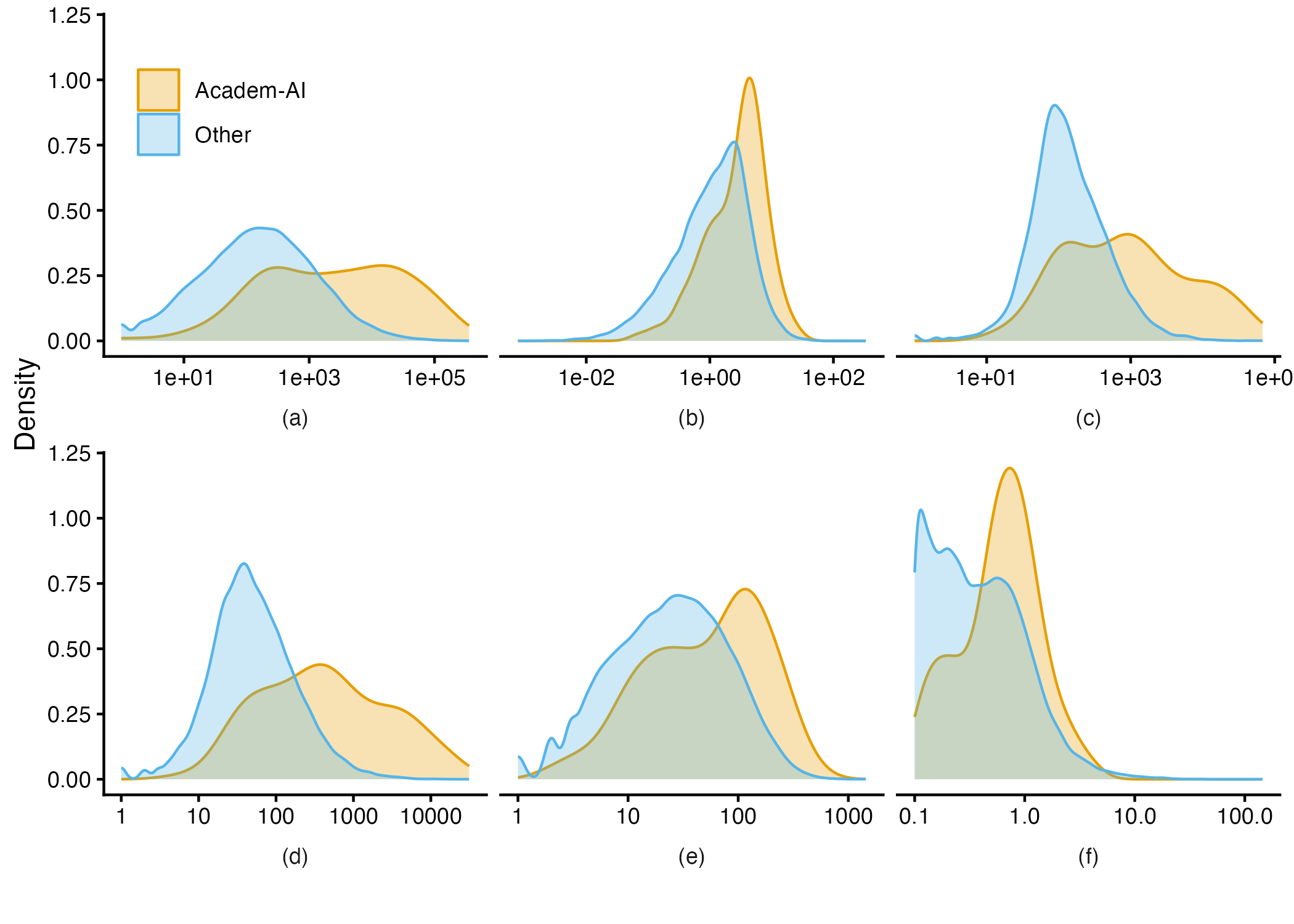}
\caption{Distributions of SJR metrics for journals stratified by whether
or not they are represented in the Academ-AI dataset; logarithmic scale;
\(P < 0.001\) for all comparisons: (a) 2024 citations to documents
published 2021--23; (b) Citations per document published 2021--2023; (c)
Documents published 2021--23; (d) Documents published 2024; (e) h-index;
(f) SJR.}\label{fig-sjr}

\figrule
\end{figure*}

Of all journals, 109 (22\%) were indexed in the SJR database, publishing 139 articles in the dataset between them (22\% of all journal articles). Compared to other journals in the SJR database, journals represented on Academ-AI had significantly higher median SJR scores, h-indices, publication outputs, citations, and citation rates ($P<0.001$ for all comparisons; \figref{fig-sjr}).

\subsection{Summary of findings}\label{summary-of-findings}

There seem to be hundreds of examples of undeclared AI-generated content in the academic literature; approximately one in four of these
evaded detection by a major publisher with an explicit policy against
such content. Less than one in twenty cases is corrected post
publication, fewer are corrected formally, and fewer still are corrected
sufficiently to comply with the publishers' own guidelines. Journals
represented on Academ-AI more commonly have APCs, and their APCs and
citation scores are higher than those of their peers in the DOAJ and SJR
database, respectively; in other words, the journals committing
editorial oversights concerning AI are paradoxically paid more for
editorial processing and achieve greater academic impact.

\section{Discussion}\label{discussion}

It seems that in a remarkable number of cases, editorial and peer review
have failed to uphold publishing standards with regard to AI. Part of the problem may be a lack of awareness of the
idiosyncratic phrasing used by LLMs. While some features, such as
explicit self-identification ``as an AI language model'' are difficult
to miss, it is unsurprising that others, such as inappropriate use of
conversational phrasing or the first-person singular, do not immediately
lead reviewers to think of AI specifically. Phrasing such as ``as of my
last knowledge update'' should, however, alert editors and reviewers
that \emph{something} is amiss, and if errors such as these endure the
scrutiny of multiple reviewers and editors, what else are they
overlooking?

The Shoukat et al.\ case is particularly illustrative of
this problem.\cite{shoukat2024} The appearance of the phrase ``regenerate response'' in
the reference list of an article does not in itself diminish the
scientific quality of that article any more than a slight misspelling or
a poorly formatted paragraph break. However, as an indicator of AI
involvement, the error led to the discovery that almost a quarter of the
reference list was likely confabulated. That 18 of the authors' sources
did not exist went unnoticed throughout the peer review, copyedit, and
typesetting processes, and had the authors been fractionally more
careful when copying the chatbot's output---avoiding copying the button
label along with the response---the confabulation may never have come to
light. The tendency of LLM chatbots to confabulate references is well
documented\cite{moskatel2024}, and many citations to non-existent sources
may already be present in the published literature.

\subsection{Dark AI}\label{dark-ai}

In the examples described here, the AI-generated text is sufficiently
incongruous with the remainder of the article as to be obvious, but LLM
outputs vary widely, and identifying them amid human-authored text is
not always straightforward or even possible. Existing AI detection tools are
unreliable and likely destined to continue in an arms race with increasingly
advanced LLMs.\cite{cheng2025}

The examples on Academ-AI were predominantly identifiable due to
characteristic turns of phrase, which are only reproduced in a
manuscript when authors copy text directly from an AI system with little
to no scrutiny. Light proofreading would be sufficient to remove such
clear AI fingerprints, though not necessarily sufficient to identify
deeper AI-induced problems in the content, such as confabulated
information. In other words, should characteristic AI phraseology become
widely known or AI detectors more reliable, unscrupulous authors
may be incentivized to expend moderate additional effort disguising
AI-generated text rather than significant additional effort thoroughly
validating it; this would mask the symptoms while failing to treat the
underlying disease.

It is entirely possible that lightly edited AI-generated text is already
abundant in the literature, undetectable, but just as, if not more,
flawed than its caveat-riddled, conversationally toned, and therefore
identifiable original version. While this hypothesis is deeply
troubling, it should not deter us from studying and contending with the
examples that \emph{are} observable, any more than the opacity of dark
matter and dark energy should deter physicists from the study of
observable matter and energy.

It is imperative that we identify and address the instances of
undeclared AI that we \emph{can} detect. In such cases, journals and
publishers must enforce their own policies via correction or retraction,
depending on the severity of the policy violation. To do otherwise sets
a precedent of policy violations going unaddressed, suggesting to
authors that declaration of AI usage is unnecessary. By contrast, a
precedent of enforcing these policies incentivizes declaration among
careless and even unscrupulous authors on the off-chance that the use of
AI is detectable in their case.

I have previously drawn a comparison to conflict of
interest.\cite{glynn2025b} Briefly, scientific research has successfully
transitioned in recent decades from a prohibitive model, in which
competing interests are disqualifying, to a transparent model, in which
competing interests are disclosed and managed.\cite{iom1991} With
disclosure becoming commonplace and failure to disclose meeting severe
consequences, researchers have every incentive to disclose their
conflicts rather than concealing them. We can use the same incentives to
encourage transparency with regard to AI usage, but this depends, again, on
journals enforcing their own policies.

\subsection{Limitations}\label{limitations}

Since Academ-AI relies on manual search and examination, data may not be
fully up to date or representative of undeclared AI in the literature as
a whole. Coverage is limited to English-language literature indexed in
online databases accessible to the author, principally Google Scholar.

\subsection{Conclusion}\label{conclusion}

Given the risks posed by the inclusion of AI-generated text in research
literature, publishers and journals must establish a precedent of
enforcing their policies against undeclared AI usage in the writing
process. When policy violations that escape the notice of editors and
peer reviewers are identified post publication, formal corrections or
retractions must be made.

\section*{Declarations}

\subsection*{Funding}

The author received no specific funding for this work.

\subsection*{Conflict of interest}
The author declares no conflict of interest.

\subsection*{Data availability}

Glynn A. Academ-AI data dump. Figshare, 2025. \href{https://doi.org/10.6084/m9.figshare.29650481.v2}{10.6084/m9.figshare.29650481.v2}

\printbibliography

\csloptions{0}{class = {american-medical-association = in-text}}
\cslcitation{glynn2024c@1}{\textsuperscript{\cslcite{glynn2024c}{1}}}
\cslcitation{hosseini2023@1}{\textsuperscript{\cslcite{hosseini2023}{2}}}
\cslcitation{hosseini2023@2}{\textsuperscript{\cslcite{hosseini2023}{2}}}
\cslcitation{openai2022@1}{\textsuperscript{\cslcite{openai2022}{3}}}
\cslcitation{weston2021@1}{\textsuperscript{\cslcite{weston2021}{4}}}
\cslcitation{nist2024@1}{\textsuperscript{\cslcite{nist2024}{5}}}
\cslcitation{charlotin2025@1}{\textsuperscript{\cslcite{charlotin2025}{6}}}
\cslcitation{ouyang2022@1}{\textsuperscript{\cslcite{ouyang2022}{7}}}
\cslcitation{retractionwatch2024@1}{\textsuperscript{\cslcite{retractionwatch2024}{8}}}
\cslcitation{issn2024@1}{\textsuperscript{\cslcite{issn2024}{9}}}
\cslcitation{oxr2025@1}{\textsuperscript{\cslcite{oxr2025}{10}}}
\cslcitation{sjr@1}{\textsuperscript{\cslcite{sjr}{11}}}
\cslcitation{wattenberg2008@1}{\textsuperscript{\cslcite{wattenberg2008}{12}}}
\cslcitation{minzatu2024e@1}{\textsuperscript{\cslcite{minzatu2024e}{13}}}
\cslcitation{openai2024@1}{\textsuperscript{\cslcite{openai2024}{14}}}
\cslcitation{bader2024e@1}{\textsuperscript{\cslcite{bader2024e}{15}}}
\cslcitation{zhang2024e@1}{\textsuperscript{\cslcite{zhang2024e}{16}}}
\cslcitation{tsai2023e@1}{\textsuperscript{\cslcite{tsai2023e}{17}}}
\cslcitation{shoukat2024e@1}{\textsuperscript{\cslcite{shoukat2024e}{18}}}
\cslcitation{wu2024e@1}{\textsuperscript{\cslcite{wu2024e}{19}}}
\cslcitation{guo2024e@1}{\textsuperscript{\cslcite{guo2024e}{20}}}
\cslcitation{aquarius2025@1}{\textsuperscript{\cslcite{aquarius2025}{21}}}
\cslcitation{kanjarla2023e@1}{\textsuperscript{\cslcite{kanjarla2023e}{22}}}
\cslcitation{lazar2024e@1}{\textsuperscript{\cslcite{lazar2024e}{23}}}
\cslcitation{tarla2023e@1}{\textsuperscript{\cslcite{tarla2023e}{24}}}
\cslcitation{ams2024,medicine2024@1}{\textsuperscript{\cslcite{ams2024}{25},\cslcite{medicine2024}{26}}}
\cslcitation{abalkina2021@1}{\textsuperscript{\cslcite{abalkina2021}{27}}}
\cslcitation{shoukat2024@1}{\textsuperscript{\cslcite{shoukat2024}{28}}}
\cslcitation{moskatel2024@1}{\textsuperscript{\cslcite{moskatel2024}{29}}}
\cslcitation{cheng2025@1}{\textsuperscript{\cslcite{cheng2025}{30}}}
\cslcitation{glynn2025b@1}{\textsuperscript{\cslcite{glynn2025b}{31}}}
\cslcitation{iom1991@1}{\textsuperscript{\cslcite{iom1991}{32}}}

\csloptions{0}{entry-ids = {glynn2024c, hosseini2023, openai2022, weston2021, nist2024, charlotin2025, ouyang2022, retractionwatch2024, issn2024, oxr2025, sjr, wattenberg2008, minzatu2024e, openai2024, bader2024e, zhang2024e, tsai2023e, shoukat2024e, wu2024e, guo2024e, aquarius2025, kanjarla2023e, lazar2024e, tarla2023e, ams2024, medicine2024, abalkina2021, shoukat2024, moskatel2024, cheng2025, glynn2025b, iom1991}}
\begin{thebibliography}{index = 1, second-field-align = flush, line-spacing = 1, entry-spacing = 1, widest-label = 10.}

\bibitem[1.]{glynn2024c}
Glynn A. Suspected Undeclared Use of Artificial Intelligence in the Academic Literature: An Analysis of the Academ-AI Dataset. \textit{arXiv}. Preprint posted online November 20, 2024. doi:\href{https://doi.org/10.48550/arXiv.2411.15218}{10.48550/arXiv.2411.15218}

\bibitem[2.]{hosseini2023}
Hosseini M, Rasmussen LM, Resnik DB. Using AI to write scholarly publications. \textit{Account Res}. Published online January 25, 2023:1-9. doi:\href{https://doi.org/10.1080/08989621.2023.2168535}{10.1080/08989621.2023.2168535}

\bibitem[3.]{openai2022}
OpenAI. Introducing ChatGPT. OpenAI. November 30, 2022. Accessed March 11, 2024. \url{https://openai.com/blog/chatgpt}

\bibitem[4.]{weston2021}
Weston J, Shuster K. Blender Bot 2.0: An open source chatbot that builds long-term memory and searches the internet. The latest AI news from Meta. July 16, 2021. Accessed April 16, 2024. \url{https://ai.meta.com/blog/blender-bot-2-an-open-source-chatbot-that-builds-long-term-memory-and-searches-the-internet/}

\bibitem[5.]{nist2024}
National Institute of Standards and Technology (US). \textit{Artificial Intelligence Risk Management Framework: Generative Artificial Intelligence Profile}. National Institute of Standards and Technology (U.S.); 2024:error:  600-1. doi:\href{https://doi.org/10.6028/NIST.AI.600-1}{10.6028/NIST.AI.600-1}

\bibitem[6.]{charlotin2025}
Charlotin D. AI Hallucination Cases. Damien Charlotin. 2025. Accessed November 8, 2025. \url{https://www.damiencharlotin.com/hallucinations/}

\bibitem[7.]{ouyang2022}
Ouyang L, Wu J, Jiang X, et al. Training language models to follow instructions with human feedback. \textit{arXiv}. Preprint posted online March 4, 2022. doi:\href{https://doi.org/10.48550/arXiv.2203.02155}{10.48550/arXiv.2203.02155}

\bibitem[8.]{retractionwatch2024}
Retraction Watch. Papers and peer reviews with evidence of ChatGPT writing. Retraction Watch. March 18, 2024. Accessed July 8, 2024. \url{https://retractionwatch.com/papers-and-peer-reviews-with-evidence-of-chatgpt-writing/}

\bibitem[9.]{issn2024}
International Standard Serial Number International Centre. The ISSN Portal. The ISSN Portal. 2024. Accessed July 2, 2024. \url{https://portal.issn.org/}

\bibitem[10.]{oxr2025}
Open Exchange Rates. About the Open Exchange Rates API. Open Exchange Rates. 2025. Accessed November 7, 2024. \url{https://openexchangerates.org/about}

\bibitem[11.]{sjr}
SCImago. SJR – SCImago Journal \& Country Rank. SCImago. Accessed November 7, 2024. \url{https://www.scimagojr.com}

\bibitem[12.]{wattenberg2008}
Wattenberg M, Viégas FB. The Word Tree, an Interactive Visual Concordance. \textit{IEEE Transactions on Visualization and Computer Graphics}. 2008;14(6):1221-1228. doi:\href{https://doi.org/10.1109/TVCG.2008.172}{10.1109/TVCG.2008.172}

\bibitem[13.]{minzatu2024e}
Retraction: Research concerning the autonomy of the electric vehicles, simulated and measured, in the case of driving at the low and the medium speed, specific to the WLTC test. (IOP Conf. Ser.: Mater. Sci. Eng.1303 012003). \textit{IOP Conf Ser: Mater Sci Eng}. 2025;1303(1):012056. doi:\href{https://doi.org/10.1088/1757-899X/1303/1/012056}{10.1088/1757-899X/1303/1/012056}

\bibitem[14.]{openai2024}
OpenAI. GPT-4 Technical Report. \textit{arXiv}. Preprint posted online March 4, 2024. doi:\href{https://doi.org/10.48550/arXiv.2303.08774}{10.48550/arXiv.2303.08774}

\bibitem[15.]{bader2024e}
Bader R, Imam A, Alnees M, et al. Removal notice to “Successful management of an Iatrogenic portal vein and hepatic artery injury in a 4-month-old female patient: A case report and literature review” [Radiology Case Reports 19 (2024) 2106-2111]. \textit{Radiology Case Reports}. 2024;19(8):3598. doi:\href{https://doi.org/10.1016/j.radcr.2024.05.010}{10.1016/j.radcr.2024.05.010}

\bibitem[16.]{zhang2024e}
Zhang M, Wu L, Yang T, Zhu B, Liu Y. Retraction notice to “The three-dimensional porous mesh structure of Cu-based metal-organic-framework - Aramid cellulose separator enhances the electrochemical performance of lithium metal anode batteries” [Surfaces and Interfaces, Volume 46, March 2024, 104081]. \textit{Surfaces and Interfaces}. 2024;46:104550. doi:\href{https://doi.org/10.1016/j.surfin.2024.104550}{10.1016/j.surfin.2024.104550}

\bibitem[17.]{tsai2023e}
Tsai CY, Yeh YH, Tsai LH, Chou ECL. Correction: Tsai et al. The Efficacy of Transvaginal Ultrasound-Guided BoNT-A External Sphincter Injection in Female Patients with Underactive Bladder. Toxins 2023, 15, 199. \textit{Toxins}. 2023;15(6):399. doi:\href{https://doi.org/10.3390/toxins15060399}{10.3390/toxins15060399}

\bibitem[18.]{shoukat2024e}
The PLOS ONE Editors. Retraction: A comparative analysis of blended learning and traditional instruction: Effects on academic motivation and learning outcomes. \textit{PLoS ONE}. 2024;19(4):e0302484. doi:\href{https://doi.org/10.1371/journal.pone.0302484}{10.1371/journal.pone.0302484}

\bibitem[19.]{wu2024e}
Assessment of the efficacy of alkaline water in conjunction with conventional medication for the treatment of chronic gouty arthritis: A randomized controlled study: Retraction. \textit{Medicine}. 2024;103(28):e38913. doi:\href{https://doi.org/10.1097/MD.0000000000039085}{10.1097/MD.0000000000039085}

\bibitem[20.]{guo2024e}
Frontiers Editorial Office. Retraction: Cellular functions of spermatogonial stem cells in relation to JAK/STAT signaling pathway. \textit{Front Cell Dev Biol}. 2024;12:1386861. doi:\href{https://doi.org/10.3389/fcell.2024.1386861}{10.3389/fcell.2024.1386861}

\bibitem[21.]{aquarius2025}
Aquarius R, Schoeters F, Wise N, Glynn A, Cabanac G. The Existence of Stealth Corrections in Scientific Literature—A Threat to Scientific Integrity. \textit{Learned Publishing}. 2025;38(2):e1660. doi:\href{https://doi.org/10.1002/leap.1660}{10.1002/leap.1660}

\bibitem[22.]{kanjarla2023e}
Corrigendum. \textit{Eur J Mass Spectrom (Chichester)}. 2023;29(4):272-272. doi:\href{https://doi.org/10.1177/14690667231195424}{10.1177/14690667231195424}

\bibitem[23.]{lazar2024e}
Lazăr NN, Râpeanu G, Iticescu C. Corrigendum to “Mitigating eggplant processing waste’s environmental impact through functional food developing.” [Trends in Food Science \& Technology (2024) 104414]. \textit{Trends in Food Science \& Technology}. 2024;149:104535. doi:\href{https://doi.org/10.1016/j.tifs.2024.104535}{10.1016/j.tifs.2024.104535}

\bibitem[24.]{tarla2023e}
Retraction: Exploring new optical solutions for nonlinear Hamiltonian amplitude equation via two integration schemes (2023 Phys. Scr. 98 095218). \textit{Phys Scr}. 2023;98(10):109701. doi:\href{https://doi.org/10.1088/1402-4896/acf6b8}{10.1088/1402-4896/acf6b8}

\bibitem[25.]{ams2024}
Annals of Medicine \& Surgery. Annals of Medicine and Surgery: Instructions for Authors. Editorial Manager. 2024. Accessed March 19, 2024. \url{https://edmgr.ovid.com/amsu/accounts/ifauth.htm}

\bibitem[26.]{medicine2024}
Medicine. Instructions for Authors. Medicine. 2024. Accessed March 27, 2024. \url{https://journals.lww.com/md-journal/Pages/Instructions-for-Authors.aspx}

\bibitem[27.]{abalkina2021}
Abalkina A. Detecting a network of hijacked journals by its archive. \textit{Scientometrics}. 2021;126(8):7123-7148. doi:\href{https://doi.org/10.1007/s11192-021-04056-0}{10.1007/s11192-021-04056-0}

\bibitem[28.]{shoukat2024}
Shoukat R, Ismayil I, Huang Q, Oubibi M, Younas M, Munir R. A comparative analysis of blended learning and traditional instruction: Effects on academic motivation and learning outcomes. \textit{PLoS One}. 2024;19(3):e0298220. doi:\href{https://doi.org/10.1371/journal.pone.0298220}{10.1371/journal.pone.0298220}

\bibitem[29.]{moskatel2024}
Moskatel LS, Zhang N. Comparative prevalence and characteristics of fabricated citations in large language models in headache medicine. \textit{Headache}. 2024;64(1):93-95. doi:\href{https://doi.org/10.1111/head.14638}{10.1111/head.14638}

\bibitem[30.]{cheng2025}
Cheng PJ, Hu FY, Chen LY, Liu JY, Wu JH, Chen WL. Generative artificial intelligence in ophthalmology research writing: A comprehensive review of applications, detection tools, and ethical considerations. \textit{Taiwan Journal of Ophthalmology}. Published online October 24, 2025. doi:\href{https://doi.org/10.4103/tjo.TJO-D-25-00072}{10.4103/tjo.TJO-D-25-00072}

\bibitem[31.]{glynn2025b}
Glynn A. The case for universal artificial intelligence declaration on the precedent of conflict of interest. \textit{Accountability in Research}. 2025;32(6):1046-1047. doi:\href{https://doi.org/10.1080/08989621.2024.2345719}{10.1080/08989621.2024.2345719}

\bibitem[32.]{iom1991}
Institute of Medicine (US) Committee on Potential Conflicts of Interest in Patient Outcomes Research Teams. \textit{Patient Outcomes Research Teams: Managing Conflict of Interest}. (Donaldson MS, Capron AM, eds.). National Academies Press (US); 1991. Accessed July 11, 2024. \url{http://www.ncbi.nlm.nih.gov/books/NBK234452/}

\end{thebibliography}
\end{multicols}
  
\end{document}